\documentclass[10pt,twocolumn,letterpaper]{article}

\usepackage{cvpr}
\usepackage{times}
\usepackage{epsfig}
\usepackage{graphicx}
\usepackage{amsmath}
\usepackage{amssymb}
\usepackage{diagbox}
\usepackage{multirow}

\usepackage[breaklinks=true,bookmarks=false]{hyperref}

\cvprfinalcopy 

\def\cvprPaperID{****} 

\begin{document}

\title{Robust Invisible Hyperlinks in Physical Photographs \\
 Based on 3D Rendering Attacks }

\author{Jun Jia\\
{\tt\small jiajun0302@sjtu.edu.cn}
\and
Zhongpai Gao\\
{\tt\small gaozhongpai@sjtu.edu.cn}
\and
Kang Chen\\
{\tt\small 1712440708@st.usst.edu.cn}
\and
Menghan Hu\\
{\tt\small mhhu@ce.ecnu.edu.cn}
\and
Guangtao Zhai\\
{\tt\small zhaiguangtao@sjtu.edu.cn}
\and
Guodong Guo\\
{\tt\small guoguodong01@baidu.com}
\and
Xiaokang Yang\\
{\tt\small xkyang@sjtu.edu.cn}
}

\maketitle

\begin{abstract}
   In the era of multimedia and Internet, people are eager to obtain information from offline to online. Quick Response (QR) codes and digital watermarks help us access information quickly. However, QR codes look ugly and invisible watermarks can be easily broken in physical photographs. Therefore, this paper proposes a novel method to embed hyperlinks into natural images, making the hyperlinks invisible for human eyes but detectable for mobile devices. Our method is an end-to-end neural network with an encoder to hide information and a decoder to recover information. From original images to physical photographs, camera imaging process will introduce a series of distortion such as noise, blur, and light. To train a robust decoder against the physical distortion from the real world, a distortion network based on 3D rendering is inserted between the encoder and the decoder to simulate the camera imaging process. Besides, in order to maintain the visual attraction of the image with hyperlinks, we propose a loss function based on just noticeable difference (JND) to supervise the training of encoder. Experimental results show that our approach outperforms the previous method in both simulated and real situations.
\end{abstract}

\section{Introduction}

\begin{figure}[t]
\begin{center}
\includegraphics[scale=0.28]{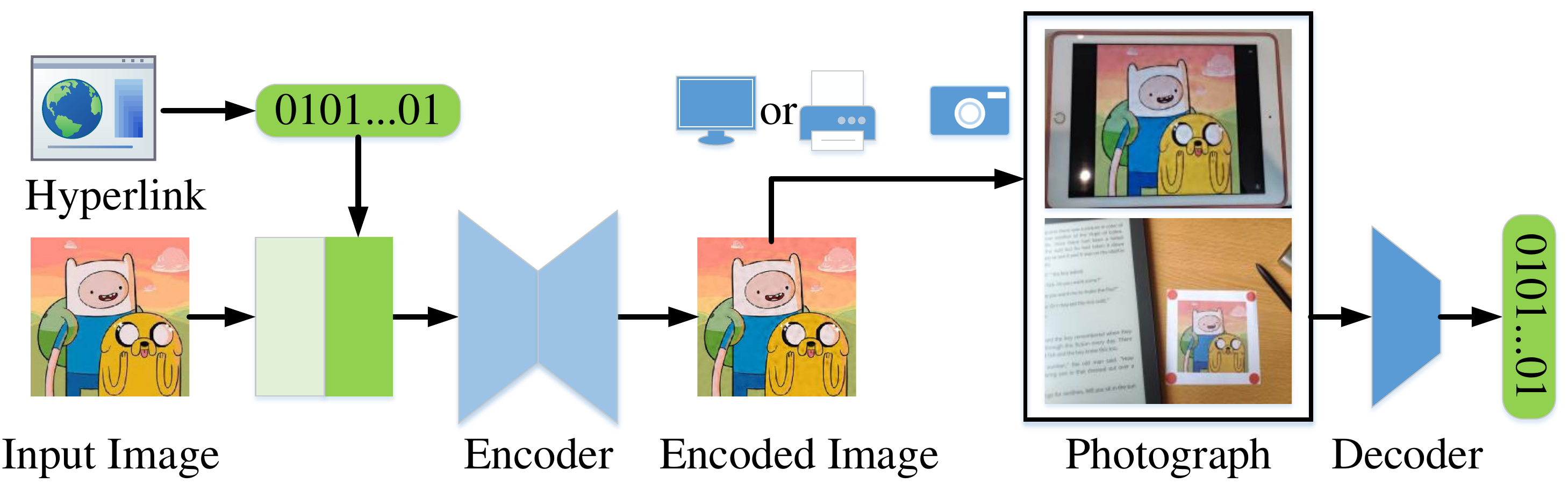}
\end{center}
\caption{Application pipeline of the proposed approach.}
\label{fig:story}
\end{figure}

Numerous technologies are emerging to build an information bridge connecting the digital world and our physical world. Quick Response (QR) codes and digital watermarks are two commonly used technologies to embed hyperlinks into images. However, QR codes not only take up space but also look ugly. Digital watermarks can be divided into visible and invisible categories. Visible watermarks have the same shortcomings as QR codes. The existing invisible watermarks are easily broken in physical photographs. People increasingly expect to add invisible information (hyperlinks) among visible information (images), thus extending the information dimension of the image to meet special demands such as advertisement and copyright protection. Hence, it is desperately to develop a state-of-the-art information hiding technique to yield images containing hyperlinks with good visual perception and allow the hidden information to be detected by mobile devices under various unconstrained environments. To meet the above requirements, this paper proposes a method to hide invisible hyperlinks into natural images. The application pipeline of this paper is shown in Figure \ref{fig:story}.

Maintaining good visual perception while ensuring the model robustness is necessary to our work. In practical applications, the camera imaging process inevitably introduces a series of distortions into original images, which will break the hidden information in images. Gao et al. \cite{TPVM} exploited the difference between human eyes and semiconductor imaging sensors in the temporal convolution of optical signals to make QR codes invisible for human eyes but detectable for mobile devices. However, this approach can be only applied to display screens and not to printed materials. Tancik et al. invented StegaStamp \cite{stegastamp} based on neural network, which hides hyperlinks into natural images and makes hyperlinks detectable for cameras. StegaStamp trains an encoder to embed hyperlinks and a decoder to recover the information. Between the encoder and the decoder, they use a distortion network to attack the output image of the encoder and send the distorted image to the decoder. StegaStamp performs well in recovering information from physical photographs by simulating the camera imaging process using the distortion network.

However, StegaStamp simplifies the camera imaging process into a series of image processing methods, which may not work in many physical situations with various lightings and shadows. In addition, it does not consider the characteristics of Human Vision System (HVS) when training the encoder. Thus, people can easily notice hidden information in images. Compared with StegaStamp, this paper exploits 3D rendering to simulate the camera imaging process in the real world: lighting, perspective warp, noise, blur, and photo compression of cameras. Experimental results show that our approach outperforms StegaStamp in many physical scenes. To further improve the visual quality of generated images, we consider the masking effect of HVS and therefore design a just noticeable difference (JND) based loss function to guide the training procedure of the encoder. The main contributions of this paper are summarized below:
\vspace{0.5em}
\begin{itemize}
\item We propose a distortion network based on 3D rendering to improve the robustness of our pipeline. Using the 3D rendering, the proposed distortion network can simulate almost all distortions from the physical world. To the best of our knowledge, our approach combines the pipeline information hiding with 3D rendering for the first time.
\item We propose a JND based loss function. Under the guidance of this loss function, the generated images with hidden hyperlinks are more visually appealing than previous work.
\item We propose a robust invisible hyperlinks approach based on the previous two points. Our approach can generate images of good visual experience containing invisible hyperlinks that are detectable by cameras on mobile devices under various unconstrained environments.
\end{itemize}

\section{Related Work}
\noindent
{\bfseries Image information hiding methods based on Generative Adversarial Networks (GANs)} With the development of neural networks, many methods in this field were proposed based on generative models. These methods are divided into three categories \cite{SteGAN_review}: cover modification, cover selection, and cover synthesis.

The cover modification includes three strategies \cite{SteGAN_review}. The first strategy is generating natural cover images by GANs, such as SGAN \cite{SGAN}, SSGAN \cite{SSGAN}, and GNCNN \cite{GNCNN}. The second strategy generates the modification probability matrix by learning distortion functions, such as ASDL-GAN \cite{ASDL-GAN} and UT-SCA-GAN \cite{UT-SCA-GAN}. The third strategy embeds the information as adversarial samples, such as the work of Tang et al. \cite{CNNAE}. The cover selection builds a mapping from cover images to hidden information, such as the work of Ke et al. \cite{Ke2019}. The cover synthesis combines cover images and information, and then generates the images with watermarks directly, such as the work of Hu et al. \cite{Hu}, SteganoGAN \cite{SteganoGAN}, HiDDeN \cite{HiDDeN}, and StegaStamp \cite{stegastamp}. In particular, HiDDeN \cite{HiDDeN} proposed a noise layer between image reconstruction and message recovery, which improves robustness against digital noise. StegaStamp \cite{stegastamp} first introduced physical noise learning in its pipeline, which is the closest paper to ours.

\vspace{0.4mm}
\noindent
{\bfseries Adversarial attacks} Adversarial attacks are proposed to add distortion to images and induce image classification or object detection networks to produce incorrect predictions. A large amount of work about adversarial examples \cite{AAAI,JPEG,Adversarial_Examples,Physical_Attacks1,Physical_Attacks2,Physical_Attacks3,Physical_Attacks4,Image_Transformation} inspires us. These papers simulate the physical imaging process and teach applications of computer vision how to resist realistic distortion.

\vspace{0.4mm}
\noindent
{\bfseries Image quality assessment (IQA) and perceptual loss} To get good quality for the generated images, we need a reasonable IQA strategy as loss functions to supervise the image reconstruction. Classic per-pixel metrics, such as Mean Squared Error (MSE) and Peak Signal-to-Noise Ratio (PSNR), are computed for each pixel independently, which cannot present structural information and perceptual quality of images. To solve this problem, several methods to measure image structural or perceptual similarity were proposed, such as SSIM \cite{SSIM}, MSSIM \cite{MSSIM}, FSIM \cite{FSIM}. In recent years, the focus of measuring the similarity between images is transferred to deep feature space, such as perceptual similarity \cite{Perceptual_Similarity}, perceptual loss \cite{Perceptual_Losses}, and LPIPS \cite{LPIPS}. These papers compute a distance from distorted images to original images in deep feature space. In addition, just noticeable difference (JND) is an important characteristic of HVS and it presents the least distortion that human can notice. Numerous JND based papers were proposed and applied to image and video compression \cite{JND1,JND2,JND3,JND4,JND5}. This paper proposes a novel loss function based on JND, which conforms to the perceptual feature of HVS. This JND-based loss is combined with perceptual similarity to guide the training process of the encoder, guaranteeing the good perceived quality of the reconstructed image.

\begin{figure*}[ht]
\begin{center}
\includegraphics[scale=0.38]{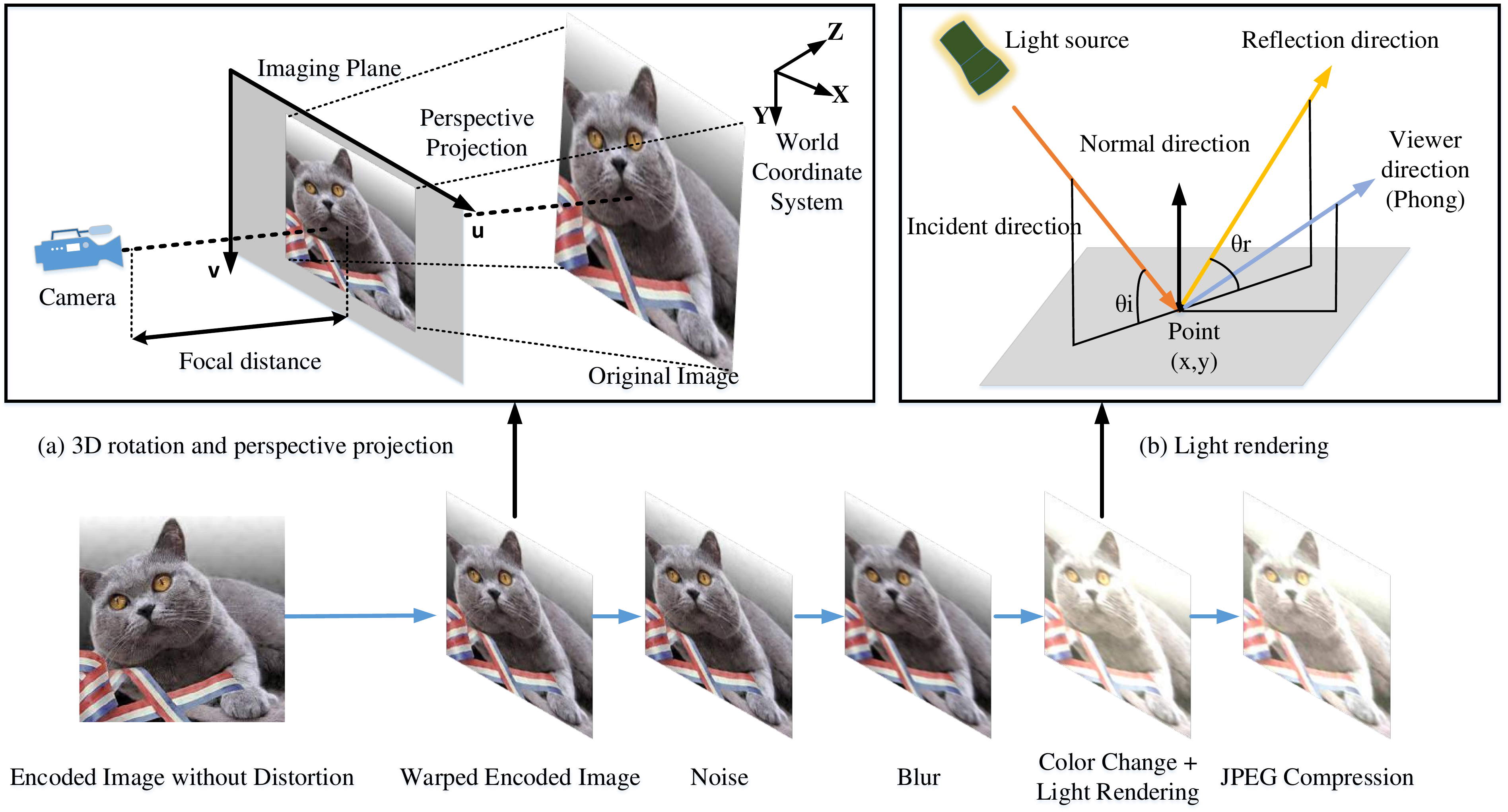}
\end{center}
\caption{The architecture of our distortion network.}
\label{fig:distortion_network}
\vspace{-1em}
\end{figure*}

\section{Distortion Network Based on Differentiable 3D Rendering}\label{sec3}

To teach our model how to resist physical distortion introduced by camera imaging, we simulate the imaging process. During training, we embed these simulation operations as a distortion network into the pipeline to process the output of the encoder and feed the processed output into the decoder. The distortion network is shown in Figure \ref{fig:distortion_network}. The idea of 3D rendering can help us simulate the camera imaging process. In order to train our model by back propagation, these operations need differentiable implements.

\subsection{3D Rotation and Perspective Projection}
Imagining a scene where a person takes photos with a pinhole camera, the len plane of the pinhole camera may not parallel to the photo plane due to the camera angle, which causes perspective warp for the photo. During training, we need to generate a random transformation matrix for each encoded image to simulate perspective warp. We simplify this problem by fixing the camera position and rotating the image plane randomly in three degrees of freedom ($x$, $y$, and $z$ axes). First, we generate three Euler angles $\alpha$, $\beta$, and $\gamma$ ($\alpha$ for $x$ axis, $\beta$ for $y$ axis, and $\gamma$ for $z$ axis) randomly. Second, we generate a 3D rotation matrix according to the Euler angles and calculate the four corner coordinates after 3D rotation. Third, we project the four corner coordinates from the 3D space to a 2D plane and generate a perspective matrix according to the coordinates before and after rotation. Finally, we use bilinear interpolation to resample the original image according to the perspective matrix.

\vspace{-0.5mm}
\subsection{Random Noise}
\vspace{-0.5mm}
The camera systems can introduce many kinds of noise during imaging \cite{Cameranoise}. We apply random Gaussian noise ($\sigma\sim U [0,0.02]$) to simulate the possible noise caused by cameras.

\vspace{-0.5mm}
\subsection{Blur}
\vspace{-0.5mm}
Blur is a common physical phenomenon in camera imaging, resulting from camera motion or defocus. To simulate motion blur, we use a random angle ranging from 0 to $2\pi$ and a blur kernel with a width ranging from 3 to 7 pixels, which is set to the same as \cite{stegastamp}. To simulate defocus blur, we apply random Gaussian noise with a width ranging from 3 to 7 pixels and a variance ranging from 0.01 to 1.

\subsection{Color Processing}
Contrast change is an inevitable distortion in camera imaging. There are many reasons for contrast change, including light, printing material, and other unknown environmental factors. Light reflection on surfaces will be described in subsection \ref{sec3.5}. In this subsection, we use the global contrast adjustment and the brightness adjustment to simulate unknown illumination factors.

\begin{figure*}[ht]
\begin{center}
\includegraphics[scale=0.38]{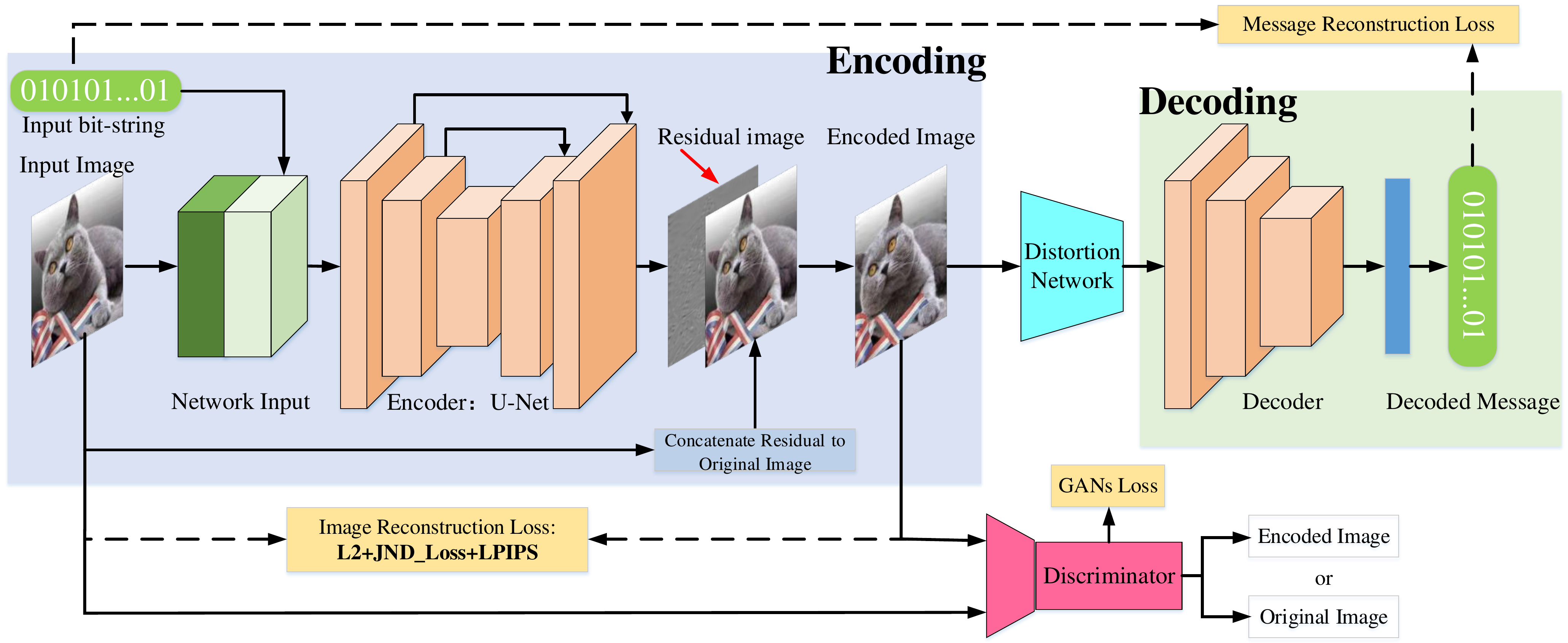}
\end{center}
\caption{The architecture of the end-to-end pipeline.}
\label{fig:framework}
\end{figure*}

\subsection{Light Reflection Rendering}\label{sec3.5}
Our ability to see the world depends on the reflection of light from the surface of any object. According to the relationship between incidence direction and reflection direction, reflection can be divided into two categories: diffuse reflection and specular reflection. The reflection type of a specific surface depends on the properties of materials. We use reflectance $\lambda$ to represent the weight of diffuse and $(1-\lambda)$ to represent the weight of specular reflection.

During training, we assume a spot light source pointing vertically to the image plane. We use a Lambertian model \cite{Lambertian} to simulate diffuse reflection on the image surface, which is defined as follows:
\begin{align}
   I_{D} &= CI_{L}\widehat{L_{n}}\cdot\widehat{N_{n}}, \notag  \\
   \widehat{L_{n}}\cdot\widehat{N_{n}} &= |N_{n}||L_{n}|\cos\theta, \tag{1}
   \label{eq1}
\end{align} where $I_{D}$ is the radiant intensity of diffuse surface, $n$ represents the $n^{th}$ points of the image, $L_{n}$ is the incident intensity, $\widehat{L_{n}}$ is the incident vector, $\widehat{N_{n}}$ is the surface normal vector, and $\theta$ is the incident angle \cite{Lambertian_Reflectance}. We use a Phong model \cite{Phong} to simulate the specular reflection as follows:
\begin{align}
I_{S}= CI_{L}(\widehat{R_{n}}\cdot\widehat{V_{n}})^s, \tag{2}
\label{eq2}
\end{align} where $\widehat{R_{n}}$ is the vector with ideal reflected direction and $\widehat{V_{n}}$ is the vector pointing to viewers. Finally, we weight the ambient light, specular radiance, and diffuse radiance to get the total light intensity:
\begin{align}
I_{T}= I_{A}+\lambda I_{D}+(1-\lambda)I_{S}, \tag{3}
\vspace{-0.5em}
\label{eq3}
\end{align} where $I_{A}$ is the global brightness adjustment of the image, presenting the ambient light.
The vectors in equation (\ref{eq1},\ref{eq2}) are normalized by $L_{2}$ function.

\vspace{-0.3em}
\subsection{JPEG Compression}
Cameras usually save images in compressed formats, which can cause quality degradation. JPEG compression is the most common compressed format, which compresses images in the transform domain. In addition to the above distortion categories, we add JPEG compression to the end of our distortion network.

\vspace{-0.5em}
\section{Model}

Figure \ref{fig:framework} presents the architecture of our approach, which is an end-to-end pipeline. The pipeline can be divided into four modules: encoder, discriminator, distortion network, and decoder. The encoder reconstructs images with original images and hidden strings in them with the help of discriminator, which is based on the idea of GANs. The distortion network mentioned in section \ref{sec3} is used to attack the output of the encoder. The decoder receives these attacked images and is trained to recover the hidden strings. The design idea of the distortion network and decoder is from adversarial attacks.

\subsection{Encoder}
The goal of the encoder is to reconstruct a new image with an original image and a hidden string, ensuring good quality for the reconstructed image. We select U-Net \cite{U-net} architecture as our encoder. The inputs are a random RGB image with a size of 400$\times$400 image and a random binary bit string with a length of 100 bits representing a hyperlink. To accelerate convergence during training, we adopt the input presentation skill of \cite{HiDDeN}: the string is reshaped to the same size as the input image and concatenated behind the image channels. Thus, a tensor with a size of 400$\times$400$\times$6 is fed to the encoder. The encoder outputs an RGB residual image presenting the hidden string. The encoded image is a combination of the residual image and the original image.

To get good quality for the encoded image, we minimize a joint loss function to supervise the encoder training. First, we use $L_{2}$ loss to calculate pixel distances in YUV space. Second, we use LPIPS to evaluate the perceptual difference between the encoded image and the original image. Third, we propose a novel loss function based on JND to supervise the location where the message is hidden. JND is the least distortion that can be noticed by HVS. Considering the view of the original image, the embedded message is noise and should be hidden in imperceptible positions. First, we generate a JND map of RGB channels for each cover image as ground truth. Then, we normalize the residual map and calculate the distance between the residual map and the JND map. We select the JND model proposed by Wu et al. \cite{JND5} to generate our ground truth. The loss function is defined as follows:
\vspace{-0.3em}
\begin{align}
L_{jnd}= L_{2}(\sigma M_{jnd}-|M_{r}|), \tag{4}
\vspace{-0.5em}
\label{eq4}
\end{align} where $\sigma$ is used to control the noise level and is set to 0.1 in this paper. With the guidance of this JND-based loss function, the encoded image gets better perceptual scores than the one without its guidance.

\subsection{Distortion Network}
We describe the details of our distortion network in section \ref{sec3} where all the simulations except for 3D rotation and perspective projection are implemented to be differentiable. The differentiable requirement ensures that the derivative is nonzero when the pipeline is trained by back propagation and gradient descent. The rounding operation of JPEG compression has zero derivatives near zero, so we use a differentiable implementation proposed by \cite{JPEG}. We implement differentiable light rendering operations with the help of Tensorflow Graphics \cite{TensorflowGraphicsIO2019}. 3D rotation and perspective projection only generate transformation matrices for distortion network instead of attacking encoded images during training, thus these two operations need not be differentiable.

\subsection{Decoder}
The decoder receives the encoded images under distortion attacks and is trained to retrieve the hidden string. Similar to StegaStamp \cite{stegastamp}, we use a spatial transformer network (STN) \cite{STN} to resist slight perspective distortion caused by camera angles. Our decoder consists of a series of convolutional layers, following by a dense layer with the same length as the input message. \textit{Sigmoid} function is used to activate the dense layer, outputting the decoded string. We use the cross entropy to supervise the training of our decoder.

\subsection{Discriminator}
To improve the quality of encoded images, we use the discriminator of WGAN-style \cite{Wasserstein} to supervise the training of our encoder.

\subsection{Loss Function}
The loss function we used during training can be divided into three parts: (\textit{i}) image reconstruction loss for encoder training: $L_{2}\ loss$, \textit{LPIPS} ($L_P$), and JND based loss ($L_{jnd}$), (\textit{ii}) discriminator loss: Wasserstein distance ($L_{w}$), and (\textit{iii}) message reconstruction loss for decoder training: cross entropy ($L_{d}$). The joint loss function of our approach is defined as follows:
\vspace{-0.5em}
\begin{align*}
L_{total}= \lambda_{1}L_2+\lambda_{2}L_{jnd}+\lambda_{3}L_P+\lambda_{4}L_d+\lambda_{5}L_w,\tag{5}
\vspace{-0.5em}
\label{eq5}
\end{align*}
During training, the loss function weight $\lambda$ increases linearly from zero. To make the decoder adapt to the distortion attacks gradually, the random change range of distortion degree increases linearly from zero.
\section{Experimental Results}

In this section, we firstly validate the design of our modules by ablation tests. Secondly, we compare our work with the closest paper, StegaStamp \cite{stegastamp}. Thirdly, we discuss the information capacity of our approach by changing the lengths of hidden strings. Finally, we test our approach by capturing photos in the real world. We select VOC2012 as the cover image dataset during training and download 100 images not contained in training set as our test set. We resize both training images and test images to 400$\times$400 resolution.

\subsection{Ablation Study}
\begin{figure}[ht]
\begin{center}
\includegraphics[scale=0.33]{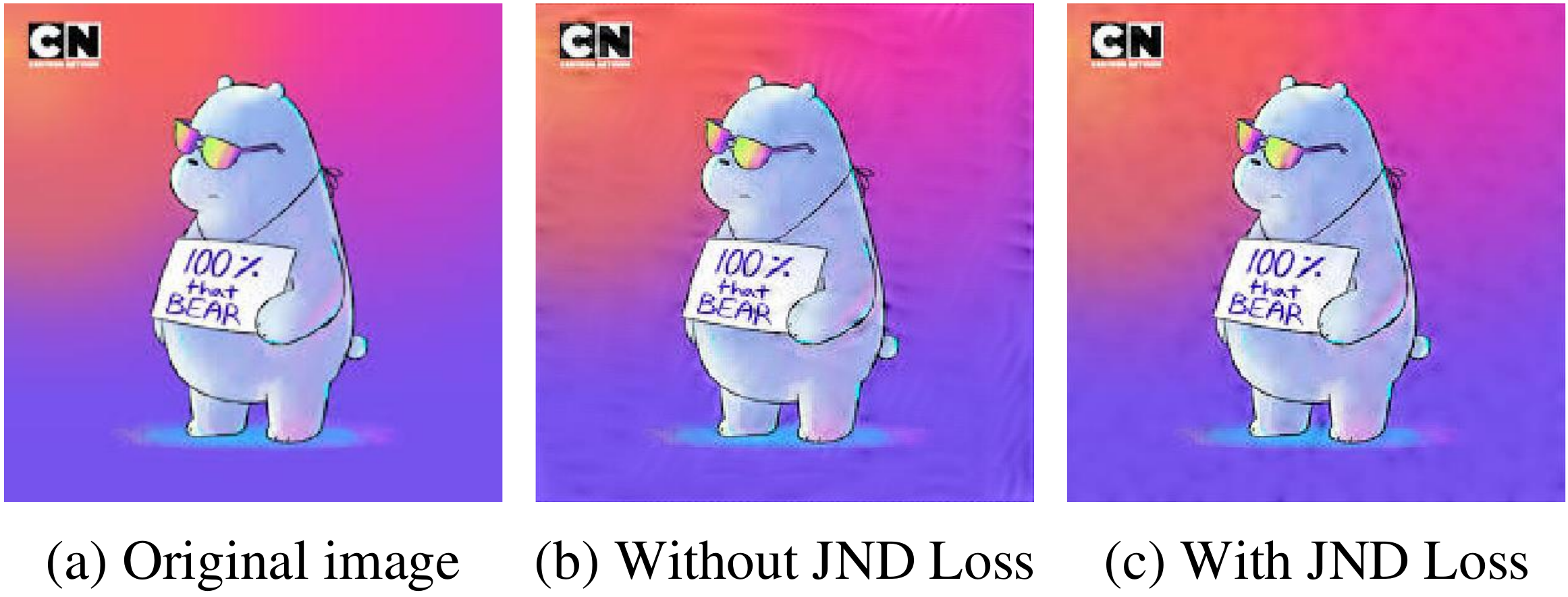}
\end{center}
\caption{The effect of the JND based loss function. Under the guidance of JND, the information is embedded in the regions with rich textures. These regions have the high masking effect to HVS.}
\label{fig:JND}
\end{figure}

First, we present the results of the ablation test for the JND-based loss function. In experiments, we train a model without the guidance of JND based loss, maintaining other modules consistent with the complete model. Figure \ref{fig:JND} presents the effect of the JND based loss function to reconstructed images, showing the information is embedded in the regions with rich textures under the guidance of JND. These regions have a high masking effect to HVS therefore are not easily noticed by human eyes.

\begin{figure}[ht]
\begin{center}
\includegraphics[scale=0.38]{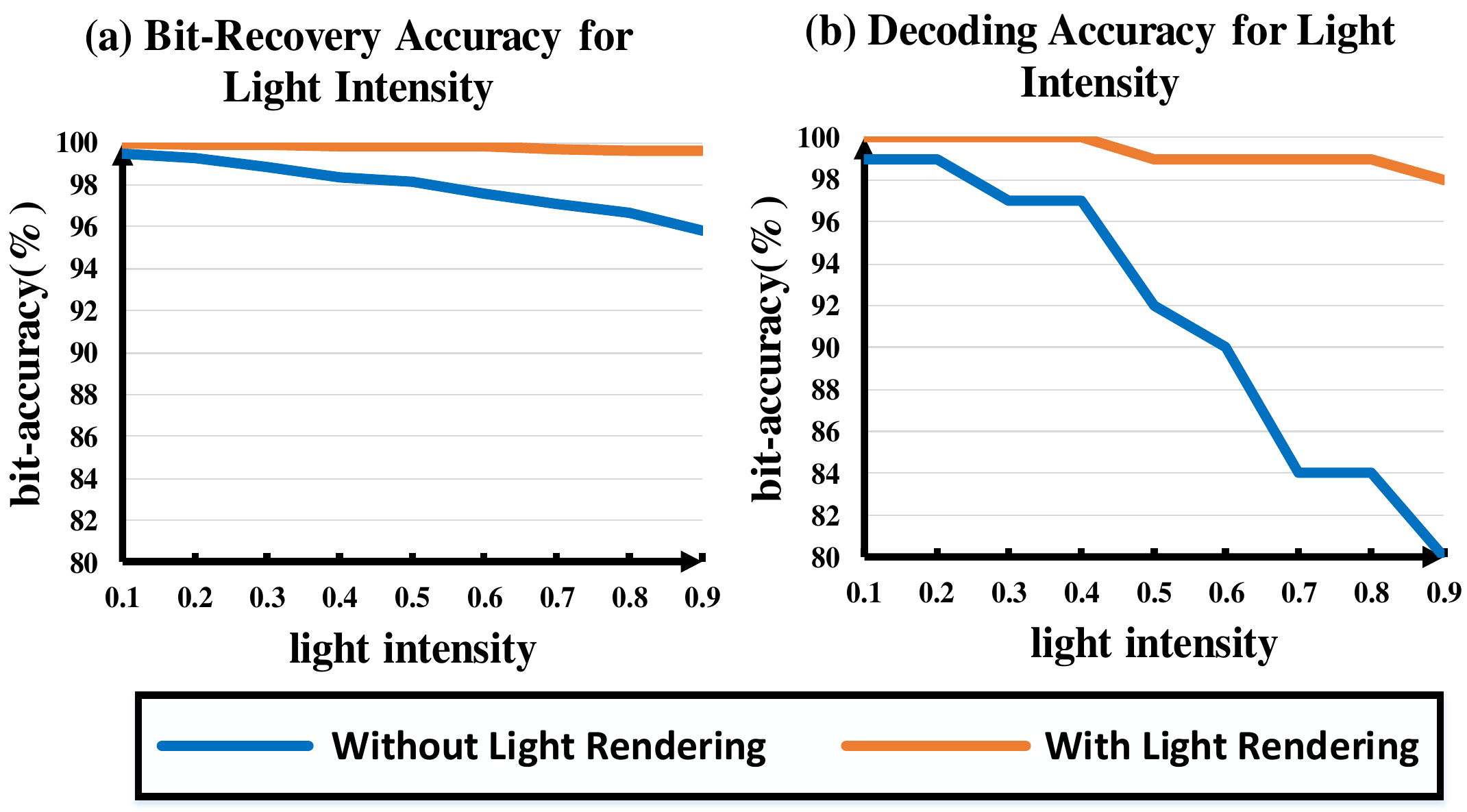}
\end{center}
\caption{Alation results for light rendering. The network trained with rendering is more robust against light reflection perturbation.}
\label{fig:ablation}
\end{figure}

\begin{figure*}[ht]
\begin{center}
\includegraphics[scale=0.38]{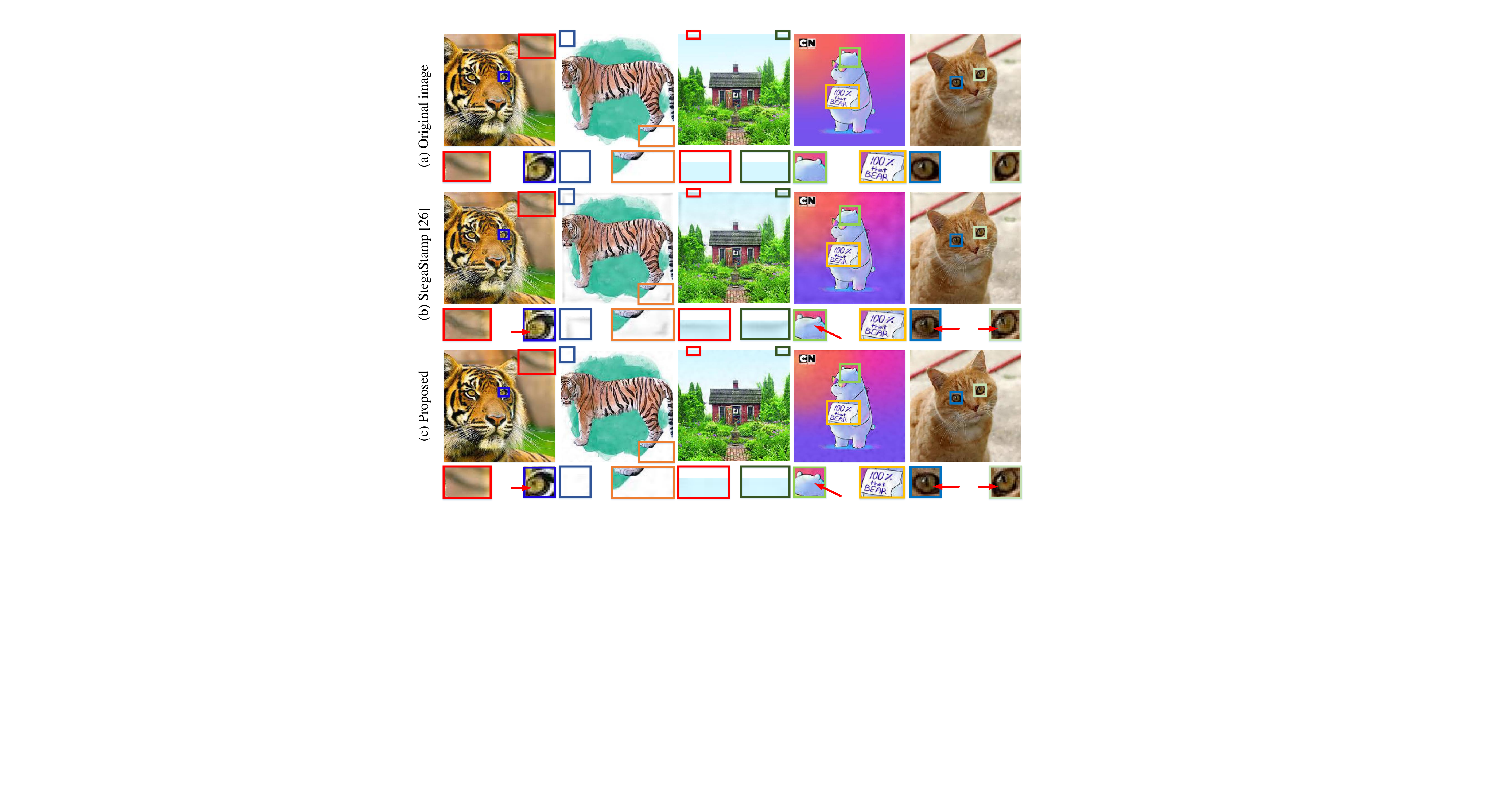}
\end{center}
\caption{The comparison results of image quality between StegaStamp \cite{stegastamp} and the proposed approach. Because a JND-based loss function is used to guide the embedding of information, the proposed approach hides information in the positions which are not easily noticed by the human eyes and get better perception quality. Colorful bounding boxes show more image details.}
\label{fig:comparison}
\vspace{-1em}
\end{figure*}

Second, to validate the effect of light rendering, we train a model without the light rendering in the distortion network, maintaining other modules unchanged. We select bit-recovery accuracy and decoding accuracy as metrics. The decoding accuracy is the decoding success rate after BCH \cite{BCH} error correcting code encoding. Figure \ref{fig:ablation} shows that light rendering does improve the robustness of hidden information against light reflection perturbations.

\subsection{Comparison Evaluation}\label{sec5.2}
Among the previous work, StegaStamp \cite{stegastamp} is the only one to perform robustness against physical distortion. Thus, we compare the performance of the proposed approach with StegaStamp \cite{stegastamp}.

\noindent
{\bfseries Metrics:} We evaluate both image reconstruction quality and robustness of information recovery. We select SSIM \cite{SSIM} and PSNR as metrics to evaluate the quality of encoded images. The bit-recovery accuracy and the decoding accuracy are applied to evaluate model robustness.

The comparison results of image reconstruction quality are shown in Table \ref{table:quality}. Table \ref{table:quality} shows that our approach gets a higher score on both SSIM and PSNR than StegaStamp \cite{stegastamp}, suggesting that our approach is more visually appealing to HVS than it. Figure \ref{fig:comparison} further verifies these results. Figure \ref{fig:comparison} shows that our approach hides the information in the positions that are not easily noticed by HVS. This suggests that the JND based loss function can effectively supervise the image reconstruction.

\begin{figure*}[ht]
\begin{center}
\includegraphics[scale=0.38]{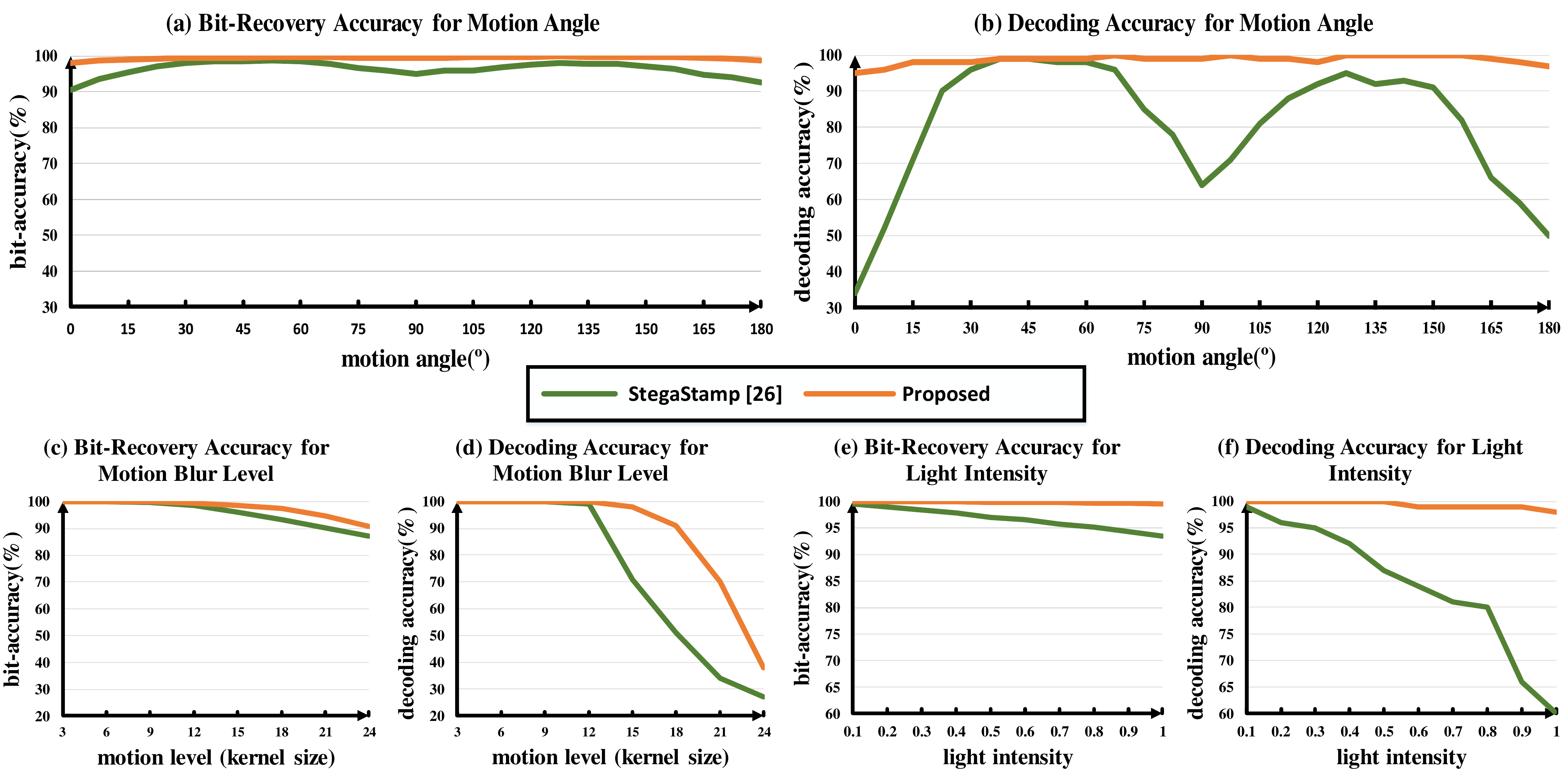}
\end{center}
\caption{Robustness evaluation results of StegaStamp \cite{stegastamp} and the proposed approach. We select motion blur, defocus blur, and light rendering to attack the encoded images. We evaluate both decoding accuracy and bit-recovery accuracy under different distortion levels. The length of the embedded message is 100 bits. When we evaluate decoding accuracy, we use BCH \cite{BCH} codes as the encoding format where 7 bytes as data codes and 5 bytes as error correcting code.}
\label{fig:plot}
\vspace{-1em}
\end{figure*}

\begin{table}[ht]
\begin{center}
\begin{tabular}{|c|c|c|}
\hline
Method & SSIM & PSNR  \\
\hline
StegaStamp \cite{stegastamp} & 0.9233 & 27.24 \\
Proposed & \textbf{0.9362} & \textbf{28.60} \\
\hline
\end{tabular}
\end{center}
\caption{The results of image reconstruction quality.}
\label{table:quality}
\vspace{-0.5em}
\end{table}

We evaluate the robustness on motion blur, defocus blur, and light reflection in simulated tests. As for camera noise and compression, they are caused by a series of complex factors depending on camera imaging environments so that we evaluate them in real environments. As for perspective warp, the influence of warp can be removed by geometric correction. These unused distortion types will be presented in section \ref{sec5.4}.

We evaluate robustness under different distortion levels as shown in Figure \ref{fig:plot}. As for motion blur, we evaluate the influence of different motion angles and blur levels (kernel size of motion blur). As for light rendering, we evaluate the influence of different light intensities. As for defocus blur, the kernel size of the Gaussian filter determines the blur level and we find both StegaStamp and the proposed approach perform well against different defocus blur levels so the results of defocus blur are not shown in Figure \ref{fig:plot}. The results show that our approach outperforms than StegaStamp in the robustness against blur and light, though StegaStamp also adds blur in its distortion network. The reason may be from the light rendering which can introduce much more corruptions to reconstructed images, improving the global robustness against other distortions.

In experiments, we find that the process of image reconstruction is most sensitive to perspective warp among these distortion categories. During training, the distortion level is set to increase linearly and the growth rate of the perspective warp is set to be the lowest. The maximum range of the 3D rotation is plus or minus two degrees around each axis, which gets a good trade off between image quality and robustness. Different from this paper, StegaStamp \cite{stegastamp} perturbs the four corner locations of the image randomly within a fixed range and generates a transformation matrix from 2D to 2D, which does not conform to the real camera imaging process.

\subsection{Capacity Evaluation}
In this subsection, we evaluate the effect of capacity on image quality. We retrain our model to have the capacity to encode 50 bits, 100 bits, and 200 bits. Table \ref{table:capacity} shows the results of image quality and robustness under different capacities. We can find that both image quality and robustness are inversely proportional to capacity. To get a good trade off between image quality, robustness, and information capacity, we select 100 bits, which is enough for hyperlinks after processed by URL-shortening tools.

\begin{table}[ht]
\begin{center}
\begin{tabular}{|c|c|c|c|}
\hline
Metric & 50 bits & 100 bits & 200 bits \\
\hline
PSNR & 29.81 & 28.60 & 28.01 \\
SSIM & 0.9474 & 0.9362 & 0.9320 \\
Bit-Accuracy & 100\% & 100\% & 96.59\% \\
\hline
\end{tabular}
\end{center}
\caption{Model performance trained with different information capacities.}
\label{table:capacity}
\end{table}

\subsection{Real Environment Evaluation}\label{sec5.4}
\begin{figure}[ht]

\begin{center}
\includegraphics[scale=0.38]{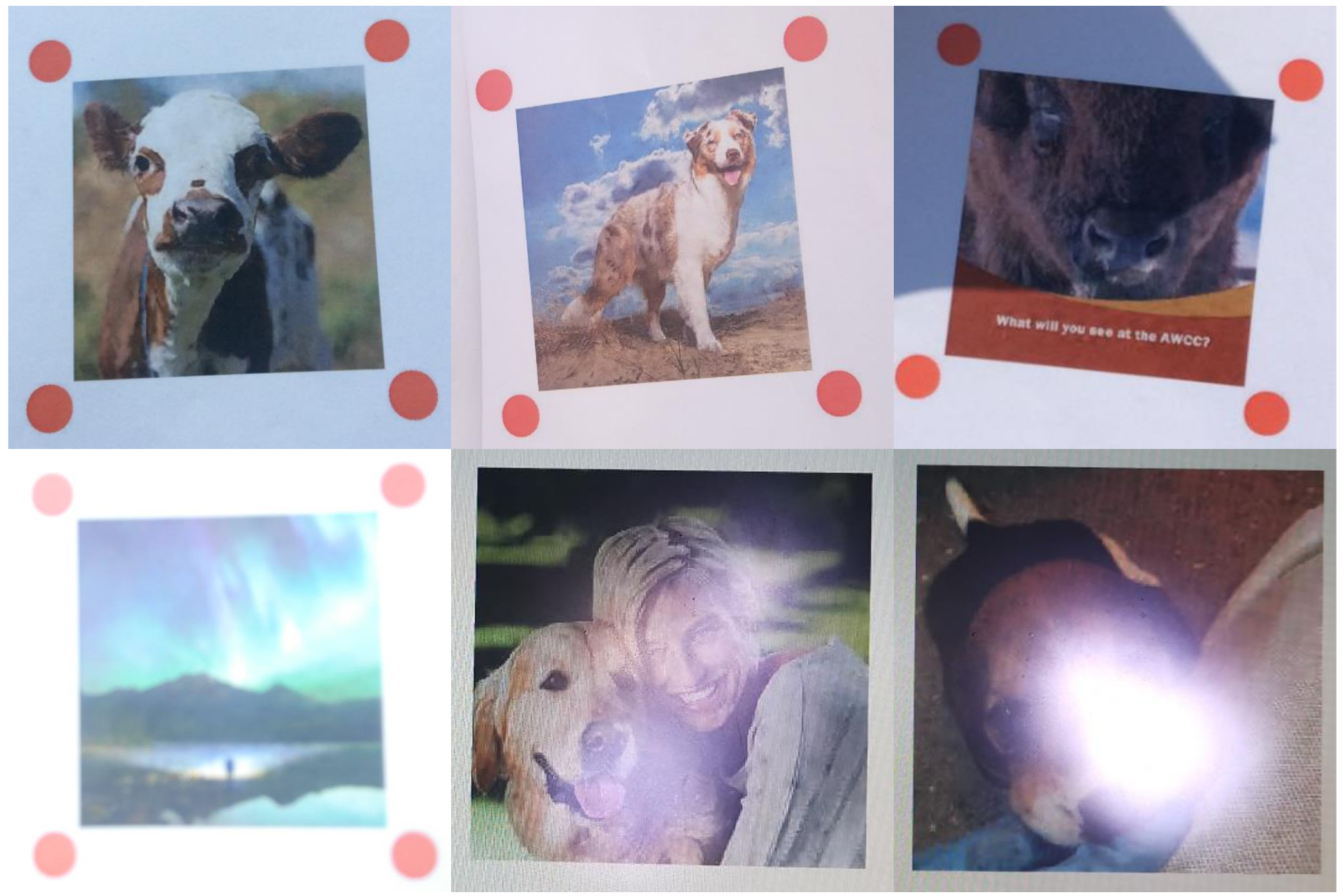}
\end{center}
\caption{Some test photos in real environments. These challenging samples are captured under different materials, different lighting conditions, and other different conditions such as blur, noise, and warp. All the six samples can be detected by our decoder.}
\label{fig:real}
\end{figure}

Finally, we evaluate our model in real environments. We use two common mobile devices to capture photographs: mobile phone camera and iPad camera. For phone cameras, we select two different phone types. Our test set includes many photos captured in challenging situations: different materials, sunlight, spot light, shadow, blur, noise, and warp, which is shown in Figure \ref{fig:real}. We locate the images with the help of four red points. For those undetected and inaccurate detected photographs, we crop and correct them by hand.

We chose different cameras in tests to introduce different noise and compress photographs in different types, which cannot be simulated well in subsection \ref{sec5.2}.  With respect to the presentation, we consider two common forms: printed on papers and showed on display screens. For the images printed on papers, the photos include indoor samples, outdoor samples, and samples with casual camera angles. For the samples displayed on screens, the photos are captured indoor, with casual camera angles, and under spot lighting conditions. We capture 20-50 photos with each camera under each condition, which is shown in Table \ref{table:real}. In addition, StegaStamp and our model respectively produce 50 images, and these generated images are captured under the same light conditions. These 100 photos can be divided into three light conditions (slight light, moderate light, and strong light) as shown in Table \ref{table:light}. Our test dataset includes 1059 photos in total. Table \ref{table:real} shows our approach is robust enough in almost all common physical environments. Table \ref{table:light} validates our approach is much more robust against light than StegaStamp \cite{stegastamp} because our training pipeline includes light rendering. The detector rate of our model is 50 frames per second on an NVIDIA 1080 GPU, which can be applied to real-time detection.

\begin{table}\renewcommand{\arraystretch}{1.0}
\begin{center}
\resizebox{8cm}{!}{
\begin{tabular}{|c|c|c|c|c|}
\hline
\multicolumn{5}{|c|}{Printed on Paper} \\
\hline
\diagbox{Device}{Scene} & Indoor & Outdoor & Warp & Mean \\
\hline
Phone\#1 & 100\% & 100\% & 72.73\% & 94.59\% \\
Phone\#2 & 100\% & 100\% & 90.91\% & 96.89\%\\
iPad & 100\% & 100\% & 77.78\% & 98.63\%\\
\hline
\multicolumn{5}{|c|}{Displayed on Screen} \\
\hline
\diagbox{Device}{Scene} & Indoor & Warp & Light & Mean \\
\hline
Phone\#1 & 100\% & 100\% & 92.86\% & 96.77\% \\
Phone\#2 & 100\% & 100\% & 89.29\% & 95.77\% \\
iPad & 100\% & 90.91\% & 86.67\% & 93.46\% \\
\hline
\end{tabular}}
\end{center}
\caption{The decoding accuracy of physical photos in real environments.}
\label{table:real}
\end{table}

\begin{table}
\begin{center}
\resizebox{8cm}{!}{
\begin{tabular}{|c|c|c|c|c|}
\hline
\diagbox{Model}{Light} & Slight & Moderate & Strong & Mean \\
\hline
StegaStamp \cite{stegastamp} & 90\% & 60\% & 15\% & 48\% \\
Proposed & 100\% & 100\% & 85\% & 94\% \\
\hline
\end{tabular}}
\end{center}
\caption{The decoding accuracy under real light conditions with different intensities.}
\label{table:light}
\vspace{-0.8em}
\end{table}


\subsection{Limitation}
There are still two limitations in the proposed approach. First, although the influence of geometrical distortion can be mininzed by the STN and the perspective transformation, we find that both the encoder and the decoder are sensitive to perspective warp. The decoder may have some probabilities to fail to detect the hyperlinks and the encoded images may get relatively low quality when the change range of perspective warp is too large. Second, when capturing screens, photos may contain moire stripes that will influence the decoding robustness of the model. In addition, different presentation materials have different reflection rates which cannot be simulated well by the proposed distortion network.

\vspace{-0.5em}

\section{Conclusion}

This paper proposes an end-to-end approach to hide invisible hyperlinks into natural images. The proposed approach uses the idea of GANs to embed information and trains a decoder based on adversarial attacks. Because our approach is applied to physical photographs and the camera imaging process will introduce a series of distortions, we insert a distortion network between the encoder and the decoder to attack the reconstructed images of the encoder. To simulate the physical distortion of photographs, the distortion network is based on 3D rendering. To obtain good visual quality, this paper proposes a novel loss function based on just noticeable difference (JND). Experimental results show that our approach is robust in various real situations and outperforms the previous work in both simulated and real situations. In the future, we plan to research how to render different material properties and use them to attack the pipeline. Our work firstly combines the information hiding pipeline with 3D rendering. We believe our work can inspire a series of interesting work in the future.

{\small
\bibliographystyle{ieee_fullname}
\bibliography{egbib}

\begin{thebibliography}{10}\itemsep=-1pt

\bibitem{Wasserstein}
Martin Arjovsky, Soumith Chintala, and L{\'e}on Bottou.
\newblock Wasserstein {G}enerative {A}dversarial {N}etworks.
\newblock In {\em Proceedings of the 34th International Conference on Machine
  Learning}, pages 214--223, 2017.

\bibitem{Adversarial_Examples}
Anish Athalye, Logan Engstrom, Andrew Ilyas, and Kevin Kwok.
\newblock Synthesizing {R}obust {A}dversarial {E}xamples.
\newblock {\em arXiv preprint arXiv:1707.07397}, 2017.

\bibitem{BCH}
R.C. Bose and D.K. Ray-Chaudhuri.
\newblock On a class of error correcting binary group codes.
\newblock {\em Information and Control}, pages 68--79, 1960.

\bibitem{Physical_Attacks4}
Shang-Tse Chen, Cory Cornelius, Jason Martin, and Duen Horng~(Polo) Chau.
\newblock Shapeshifter: {R}obust {P}hysical {A}dversarial {A}ttack on {F}aster
  {R}-{CNN} {O}bject {D}etector.
\newblock In {\em Machine Learning and Knowledge Discovery in Databases}, pages
  52--68. Springer International Publishing, 2019.

\bibitem{Perceptual_Similarity}
Alexey Dosovitskiy and Thomas Brox.
\newblock Generating {I}mages with {P}erceptual {S}imilarity {M}etrics based on
  {D}eep {N}etworks.
\newblock In {\em Advances in Neural Information Processing Systems 29}, pages
  658--666. Curran Associates, Inc., 2016.

\bibitem{Physical_Attacks2}
Kevin Eykholt, Ivan Evtimov, Earlence Fernandes, Bo Li, Amir Rahmati, Florian
  Tramer, Atul Prakash, Tadayoshi Kohno, and Dawn Song.
\newblock Physical {A}dversarial {E}xamples for {O}bject {D}etectors.
\newblock {\em arXiv preprint arXiv:1807.07769}, 2018.

\bibitem{Physical_Attacks1}
Kevin Eykholt, Ivan Evtimov, Earlence Fernandes, Bo Li, Amir Rahmati, Chaowei
  Xiao, Atul Prakash, Tadayoshi Kohno, and Dawn Song.
\newblock Robust {P}hysical-{W}orld {A}ttacks on {D}eep {L}earning {V}isual
  {C}lassification.
\newblock In {\em The IEEE Conference on Computer Vision and Pattern
  Recognition (CVPR)}, June 2018.

\bibitem{TPVM}
Zhongpai Gao, Guangtao Zhai, and Chunjia Hu.
\newblock The {I}nvisible {QR} {C}ode.
\newblock In {\em Proceedings of the 23rd ACM International Conference on
  Multimedia}, pages 1047--1050, 2015.

\bibitem{Cameranoise}
Samuel~W. Hasinoff.
\newblock Photon, {P}oisson {N}oise.
\newblock In {\em Computer Vision: A Reference Guide}, pages 608--610, 2014.

\bibitem{Hu}
Donghui Hu, Liang Wang, Wenjie Jiang, Shuli Zheng, and Bin Li.
\newblock A {N}ovel {I}mage {S}teganography {M}ethod via {D}eep {C}onvolutional
  {G}enerative {A}dversarial {N}etworks.
\newblock {\em IEEE Access}, 6:38303--38314, 2018.

\bibitem{STN}
Max Jaderberg, Karen Simonyan, Andrew Zisserman, and koray kavukcuoglu.
\newblock Spatial {T}ransformer {N}etworks.
\newblock In {\em Advances in Neural Information Processing Systems 28}, pages
  2017--2025, 2015.

\bibitem{AAAI}
Steve~T.K. Jan, Joseph Messou, Yen-Chen Lin, Jia-Bin Huang, and Gang Wang.
\newblock Connecting the {D}igital and {P}hysical {W}orld: {I}mproving the
  {R}obustness of {A}dversarial {A}ttacks.
\newblock In {\em AAAI}, Jan 2019.

\bibitem{JND1}
Yuting Jia, Weisi Lin, and Ashraf~A. Kassim.
\newblock Estimating {J}ust-{N}oticeable {D}istortion for {V}ideo.
\newblock {\em IEEE Transactions on Circuits and Systems for Video Technology},
  16(7):820--829, July 2006.

\bibitem{Perceptual_Losses}
Justin Johnson, Alexandre Alahi, and Fei-Fei Li.
\newblock Perceptual {L}osses for {R}eal-{T}ime {S}tyle {T}ransfer and
  {S}uper-{R}esolution.
\newblock In {\em The European Conference on Computer Vision (ECCV)}, pages
  694--711, 2016.

\bibitem{Ke2019}
Yan Ke, Min qing Zhang, Jia Liu, Ting ting Su, and Xiao yuan Yang.
\newblock Generative steganography with {K}erckhoffs' principle.
\newblock {\em Multimedia Tools and Applications}, 78(10):13805--13818, 2019.

\bibitem{Lambertian_Reflectance}
Sanjeev~J. Koppal.
\newblock Lambertian {R}eflectance.
\newblock In {\em Computer Vision: A Reference Guide}, pages 441--443, 2014.

\bibitem{Physical_Attacks3}
Alexey Kurakin, Ian Goodfellow, and Samy Bengio.
\newblock Adversarial examples in the physical world.
\newblock {\em arXiv preprint arXiv:1607.02533}, 2016.

\bibitem{Lambertian}
J.~H. Lambert.
\newblock In {\em Photometria {S}ive de {M}ensure de {G}radibus {L}uminis,
  {C}olorum et {U}mbrae}, 1760.

\bibitem{JND3}
Anmin Liu, Weisi Lin, Manoranjan Paul, Chenwei Deng, and Fan Zhang.
\newblock Just {N}oticeable {D}ifference for {I}mages {W}ith {D}ecomposition
  {M}odel for {S}eparating {E}dge and {T}extured {R}egions.
\newblock {\em IEEE Transactions on Circuits and Systems for Video Technology},
  20(11):1648--1652, Nov 2010.

\bibitem{SteGAN_review}
Jia Liu, Yan Ke, Yu Lei, Zhuo Zhang, Jun Li, Peng Luo, Minqing Zhang, and
  Xiaoyuan Yang.
\newblock Recent {A}dvances of {I}mage {S}teganography with {G}enerative
  {A}dversarial {N}etworks.
\newblock {\em arXiv preprint arXiv:1907.01886}, 2019.

\bibitem{Phong}
B.~T. Phong.
\newblock Illumination for {C}omputer {G}enerated {P}ictures.
\newblock {\em Commun. ACM}, (6):311--317, 1975.

\bibitem{GNCNN}
Yinlong Qian, Jing Dong, Wei Wang, and Tieniu Tan.
\newblock Deep learning for steganalysis via convolutional neural networks.
\newblock In {\em Media Watermarking, Security, and Forensics 2015}, volume
  9409, pages 171--180, 2015.

\bibitem{U-net}
Olaf Ronneberger, Philipp Fischer, and Thomas Brox.
\newblock U-{N}et: {C}onvolutional {N}etworks for {B}iomedical {I}mage
  {S}egmentation.
\newblock In {\em Medical Image Computing and Computer-Assisted Intervention --
  MICCAI 2015}, pages 234--241, 2015.

\bibitem{SSGAN}
Haichao Shi, Jing Dong, Wei Wang, Yinlong Qian, and Xiaoyu Zhang.
\newblock Ssgan: {S}ecure {S}teganography {B}ased on {G}enerative {A}dversarial
  {N}etworks.
\newblock In {\em Advances in Multimedia Information Processing -- PCM 2017},
  pages 534--544. Springer International Publishing, 2018.

\bibitem{JPEG}
Richard Shin and Dawn Song.
\newblock Jpeg-resistant {A}dversarial {I}mages.
\newblock In {\em NeurIPS Workshop on Machine Learning and Computer Security},
  2017.

\bibitem{stegastamp}
Matthew Tancik, Ben Mildenhall, and Ren Ng.
\newblock Stegastamp: {I}nvisible {H}yperlinks in {P}hysical {P}hotographs.
\newblock {\em arXiv preprint arXiv:1904.05343}, 2019.

\bibitem{CNNAE}
Weixuan Tang, Bin Li, Shunquan Tan, Mauro Barni, and Jiwu Huang.
\newblock C{NN}-{B}ased {A}dversarial {E}mbedding for {I}mage {S}teganography.
\newblock {\em IEEE Transactions on Information Forensics and Security},
  14(8):2074--2087, Aug 2019.

\bibitem{ASDL-GAN}
Weixuan Tang, Shunquan Tan, Bin Li, and Jiwu Huang.
\newblock Automatic {S}teganographic {D}istortion {L}earning {U}sing a
  {G}enerative {A}dversarial {N}etwork.
\newblock {\em IEEE Signal Processing Letters}, 24(10):1547--1551, Oct 2017.

\bibitem{Image_Transformation}
Dang~Duy Thang and Toshihiro Matsui.
\newblock Image {T}ransformation can make {N}eural {N}etworks more robust
  against {A}dversarial {E}xamples.
\newblock {\em arXiv preprint arXiv:1901.03037}, 2019.

\bibitem{TensorflowGraphicsIO2019}
Julien Valentin, Cem Keskin, Pavel Pidlypenskyi, Ameesh Makadia, Avneesh Sud,
  and Sofien Bouaziz.
\newblock Tensorflow {G}raphics: {C}omputer {G}raphics {M}eets {D}eep
  {L}earning.
\newblock 2019.

\bibitem{SGAN}
Denis Volkhonskiy, Ivan Nazarov, and Evgeny Burnaev.
\newblock Steganographic {G}enerative {A}dversarial {N}etworks.
\newblock {\em arXiv preprint arXiv:1703.05502}, 2017.

\bibitem{SSIM}
Zhou Wang, Alan~Conrad Bovik, Hamid~Rahim Sheikh, and Eero~P. Simoncelli.
\newblock Image quality assessment: from error visibility to structural
  similarity.
\newblock {\em IEEE Transactions on Image Processing}, 13(4):600--612, April
  2004.

\bibitem{MSSIM}
Zhou Wang, Eero~P. Simoncelli, and Alan~Conrad Bovik.
\newblock Multiscale structural similarity for image quality assessment.
\newblock In {\em The Thrity-Seventh Asilomar Conference on Signals, Systems
  Computers, 2003}, volume~2, pages 1398--1402, Nov 2003.

\bibitem{JND5}
Jinjian Wu, Leida Li, Weisheng Dong, Guangming Shi, Weisi Lin, and C.-C.~Jay
  Kuo.
\newblock Enhanced {J}ust {N}oticeable {D}ifference {M}odel for {I}mages with
  {P}attern {C}omplexity.
\newblock {\em IEEE Transactions on Image Processing}, 26(6):2682--2693, June
  2017.

\bibitem{JND4}
Jinjian Wu, Guangming Shi, Weisi Lin, Anmin Liu, and Fei Qi.
\newblock Just {N}oticeable {D}ifference {E}stimation for {I}mages with
  {F}ree-{E}nergy {P}rinciple.
\newblock {\em IEEE Transactions on Multimedia}, 15(7):1705--1710, Nov 2013.

\bibitem{UT-SCA-GAN}
Jianhua Yang, Kai Liu, Xiangui Kang, Edward~K. Wong, and Yun-Qing Shi.
\newblock Spatial {I}mage {S}teganography {B}ased on {G}enerative {A}dversarial
  {N}etwork.
\newblock {\em arXiv preprint arXiv:1804.07939}, 2018.

\bibitem{JND2}
X.K. Yang, W.S. Ling, Z.K. Lu, E.P. Ong, and S.S. Yao.
\newblock Just noticeable distortion model and its applications in video
  coding.
\newblock {\em Signal Processing: Image Communication}, 20:662--680, 2005.

\bibitem{SteganoGAN}
Kevin~Alex Zhang, Alfredo Cuesta-Infante, Lei Xu, and Kalyan Veeramachaneni.
\newblock Steganogan: {H}igh {C}apacity {I}mage {S}teganography with {G}ans.
\newblock {\em arXiv preprint arXiv:1901.03892}, 2019.

\bibitem{FSIM}
Lin Zhang, Lei Zhang, Xuanqin Mou, and David Zhang.
\newblock F{SIM}: {A} {F}eature {S}imilarity {I}ndex for {I}mage {Q}uality
  {A}ssessment.
\newblock {\em IEEE Transactions on Image Processing}, 20(8):2378--2386, Aug
  2011.

\bibitem{LPIPS}
Richard Zhang, Phillip Isola, Alexei~A. Efros, Eli Shechtman, and Oliver Wang.
\newblock The {U}nreasonable {E}ffectiveness of {D}eep {F}eatures as a
  {P}erceptual {M}etric.
\newblock In {\em The IEEE Conference on Computer Vision and Pattern
  Recognition (CVPR)}, June 2018.

\bibitem{HiDDeN}
Jiren Zhu, Russell Kaplan, Justin Johnson, and Fei-Fei Li.
\newblock Hi{DD}e{N}: {H}iding {D}ata with {D}eep {N}etworks.
\newblock In {\em The European Conference on Computer Vision (ECCV)}, September
  2018.

\end{thebibliography}
}

\end{document}